\documentclass[11pt,twoside]{article}
\usepackage{ihep,epsfig,amsfonts,amsmath,amssymb,amsbsy}

\begin{document}

\title{
{\Large \bf \uppercase{Numerical Spherically Symmetric Static Solution~of
the RTG~Equations Outside~the~Matter}}}
\author{{\large\bf  Anton Godizov}\thanks{E-mail:
godizov@sirius.ihep.su}\\
{\small {\it Institute for High Energy Physics, Protvino, Russia}}}
\date{}
\maketitle

\vskip-1.0cm

\begin{abstract}
There was obtained a numerical external solution for the exact system of the RTG 
equations with some natural boundary conditions in the static 
spherically symmetric case. The properties of the solution are discussed.
\end{abstract}

\vspace*{1.5cm}

The Relativistic Theory of Gravity (RTG) is the theory of the self-interacting 
massive tensor field in the Minkowski space. The exact system of the RTG field 
equations is in form of \cite{logunov}
\begin{eqnarray}
\label{rtgequ1}
&&\gamma^{\alpha\beta}D_{\alpha}D_{\beta}\phi^{\mu\nu} + 
\frac{1}{r_0^2}\phi^{\mu\nu} = \frac{16\pi G}{c^4}t^{\mu\nu}\;,\\
&& D_{\mu}\phi^{\mu\nu} = 0\;,\nonumber
\end{eqnarray}
where $\gamma^{\mu\nu}$ is the metric of the Minkowski space, $D_{\mu}$ is the 
covariant derivative in the Minkowski space, $\phi^{\mu\nu}$ is the physical 
gravitational field, $r_0\equiv\frac{\hbar}{m_gc}$, $m_g$ is the gravitational 
field mass, $\hbar$ is the Planck constant, $c$ is the speed of light in vacuum, 
$G$ is the gravitational constant and $t^{\mu\nu} = t_M^{\mu\nu} + t_g^{\mu\nu}$ 
is the energy-momentum tensor both of matter (except gravitational) fields and 
the gravitational field itself. 

Due to {\it the geometrization principle} the system (\ref{rtgequ1}) may be also 
represented as the Hilbert-Einstein type system of equations \cite{logunov} 
$$
R^{\mu\nu}-\frac{1}{2}g^{\mu\nu}R + \frac{1}{2r_0^2}\left[g^{\mu\nu}+
\left(g^{\mu\alpha}g^{\nu\beta}-\frac{1}{2}g^{\mu\nu}g^{\alpha\beta}\right)
\gamma_{\alpha\beta}\right] = \frac{16\pi G}{c^4}T^{\mu\nu}\;,
$$
$$
D_{\mu}(\sqrt{-g}g^{\mu\nu}) = 0\;,
$$
where $g^{\mu\nu}$ is the metric of the effective Riemann space 
($\sqrt{-g}g^{\mu\nu} = \sqrt{-\gamma}\phi^{\mu\nu}+
\sqrt{-\gamma}\gamma^{\mu\nu}$), 
$T^{\mu\nu}$ is the tensor of matter in the effective curved space, 
and in the static spherically symmetric case can be reduced to 
\begin{eqnarray}
\label{rtgequ}
&&\frac{dW}{dr} = y\;,\;\;\;\;\frac{dU}{dr} = U\frac{V(1 + 
\frac{1}{2r_0^2}(W^2 - r^2)) - y^2 - 
\frac{W^2}{4r_0^2}(\frac{V}{U} - 1)}{Wy}\;,\\
&&\frac{dV}{dr} = V\frac{V(1 + \frac{1}{2r_0^2}(W^2 - r^2) - \frac{4yr}{W}) + 
3y^2 - \frac{W^2}{4r_0^2}(\frac{V}{U} - 1)}{Wy}\;,\nonumber\\
&&\frac{dy}{dr} = \frac{V(1 + \frac{1}{2r_0^2}(W^2 - r^2) - 
\frac{2yr}{W}) + y^2}{W} - \frac{4\pi G}{c^2}VW\rho(r)\;,\nonumber
\end{eqnarray}
where $\rho(r)$ is the density of matter and $U(r)$, $V(r)$, $W(r)$ are the 
metric coefficients for the interval 
$$
ds^2 = U(r)(cdt)^2 - V(r)dr^2 - W^2(r)(d\Theta^2 + sin^2\Theta  d\phi^2)\;,
$$
in the effective Riemann space.

Our aim is to find non-trivial solution of (\ref{rtgequ}) which satisfies the 
following conditions 
\begin{enumerate}
\item $\lim_{r\to\infty}g^{\mu\nu} = \gamma^{\mu\nu}$,
\item $\lim_{r\to\infty}\phi^{\mu\nu} = \phi_{(\infty)}^{\mu\nu}$, where 
      $\gamma^{\alpha\beta}D_{\alpha}D_{\beta}\phi_{(\infty)}^{\mu\nu} + 
       \frac{1}{r_0^2}\phi_{(\infty)}^{\mu\nu} = 0$.
\end{enumerate}
Under such conditions the only possible behavior of the effective 
metric coefficients in the infinity is 
\begin{eqnarray}
\label{asymp}
&& U(r) = 1 - C\frac{e^{-\frac{r}{r_0}}}{r} + O(e^{-\frac{2r}{r_0}})\;,\;\;
V(r) = 1 + C\frac{e^{-\frac{r}{r_0}}}{r} + O(e^{-\frac{2r}{r_0}})\;,\\
&& W(r) = r + C\frac{e^{-\frac{r}{r_0}}}{2} + O(e^{-\frac{2r}{r_0}})\;.\nonumber
\end{eqnarray}
Here we make an assumption that the asymptotic behavior of the effective metric 
coefficients is fixed mainly by the source mass, i.e.
\begin{equation}
\label{asymp1}
C\equiv\frac{2G}{c^4}\int t^{00}(\vec r)d\vec r
\approx\int t_M^{00}(\vec r)d\vec r = \frac{2MG}{c^2} = r_g\;. 
\end{equation}
There is no gauge symmetry in system (\ref{rtgequ}) and boundary 
conditions (\ref{asymp}), (\ref{asymp1}) fix a unique solution. 

\begin{figure}
\epsfxsize=8cm\epsfysize=5cm\epsffile{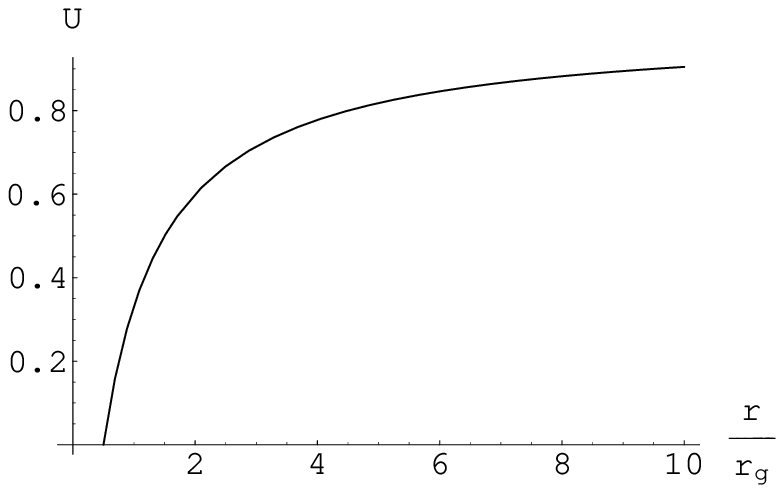}
\vskip -5cm
\hskip 9cm
\epsfxsize=8cm\epsfysize=5cm\epsffile{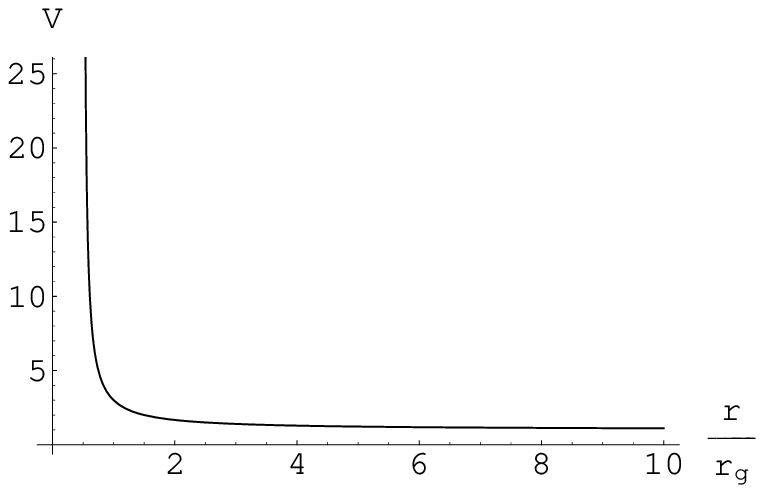}
\vskip 1.5cm
\hskip 4.5cm
\epsfxsize=8cm\epsfysize=5cm\epsffile{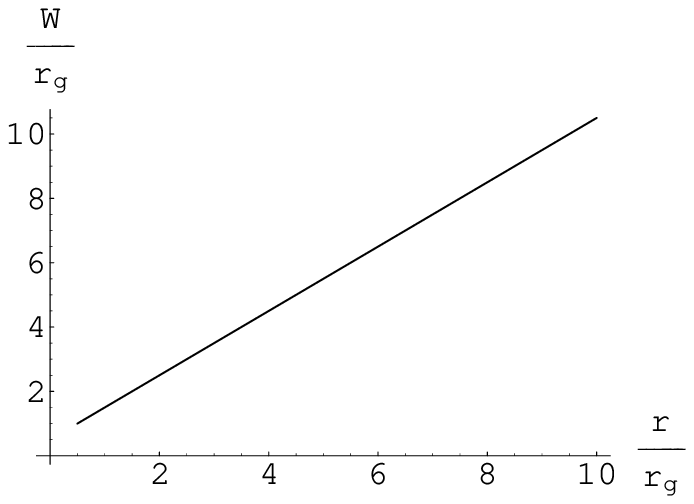}
\caption{The Schwarzschild solution in the harmonic coordinates.}
\label{garm}
\end{figure}

This solution was obtained by numerical integration in {\it Mathematica 5.0}. But 
first we did it for the system of Hilbert-Einstein equations 
in the harmonic coordinates (it can be derived by taking the limit 
$r_0\equiv\frac{\hbar}{m_gc}\to\infty$ in 
(\ref{rtgequ})) with the boundary conditions taken from the exact Schwarzschild 
solution in the harmonic coordinates 
$$
U(r) = \frac{r - \frac{r_g}{2}}{r + \frac{r_g}{2}}\;,\;\;\;
V(r) = \frac{r + \frac{r_g}{2}}{r - \frac{r_g}{2}}\;,\;\;\;
W(r) = r + \frac{r_g}{2}\;.
$$
The numerical solution for the Hilbert-Einstein equations in harmonic coordinates 
with these boundary conditions (fig. \ref{garm}) was found in the interval 
$\frac{1}{2} + 10^{-9}\equiv z_{min}\le\frac{r}{r_g}\le 10^3$. One obtains that 
$U(z_{min}r_g) = 0.9999999973\cdot 10^{-9}$, 
$V(z_{min}r_g) = 1.0000000029\cdot 10^9$, 
$U(z_{min}r_g)V(z_{min}r_g) = 1.0000000002$, $W(z_{min}r_g) = 1.000000001 r_g$.

\begin{figure}
\vskip -0.5cm
\epsfxsize=7cm\epsfysize=5cm\epsffile{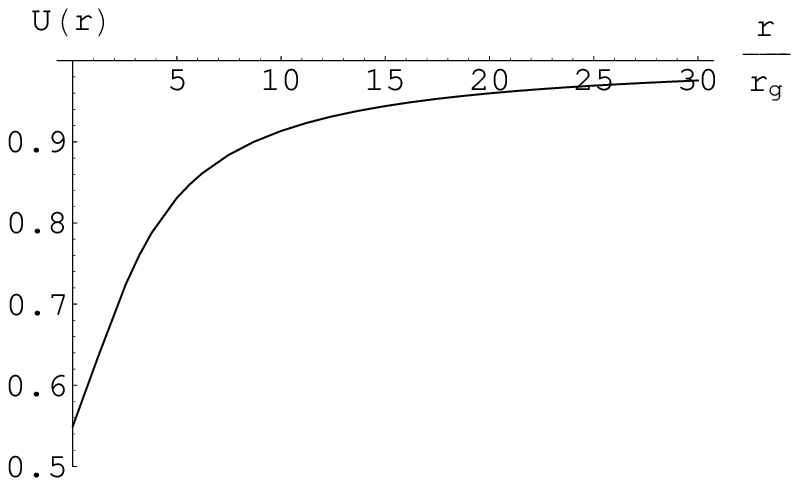}
\vskip -5cm
\hskip 9cm
\epsfxsize=7cm\epsfysize=5cm\epsffile{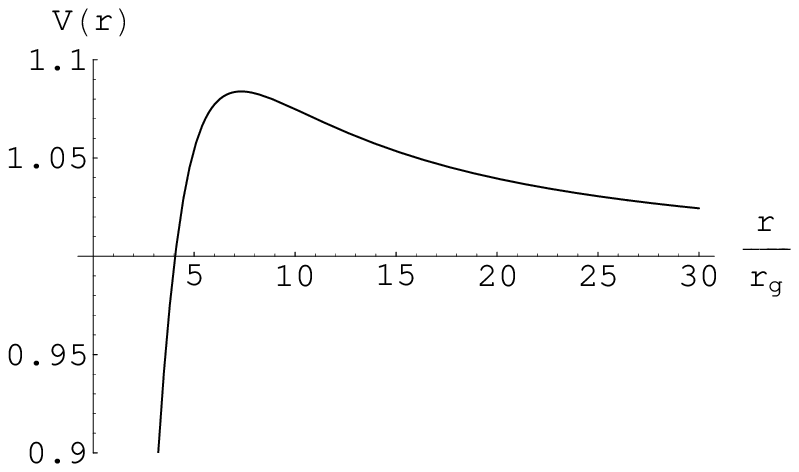}
\epsfxsize=7cm\epsfysize=5cm\epsffile{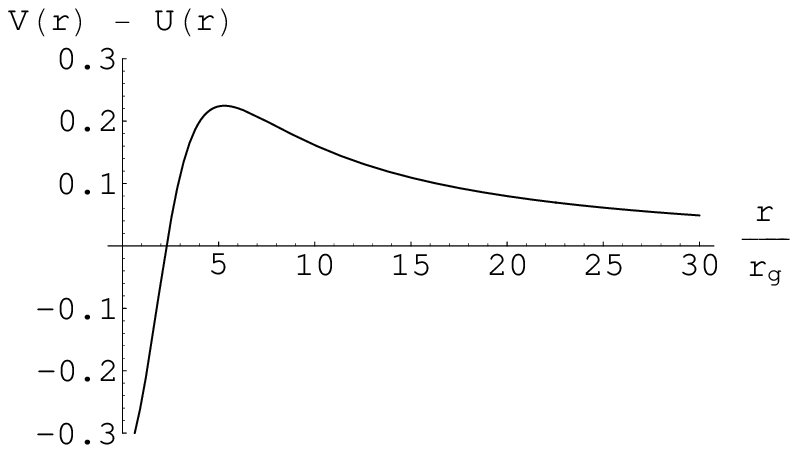}
\vskip -5cm
\hskip 9cm
\epsfxsize=7cm\epsfysize=5cm\epsffile{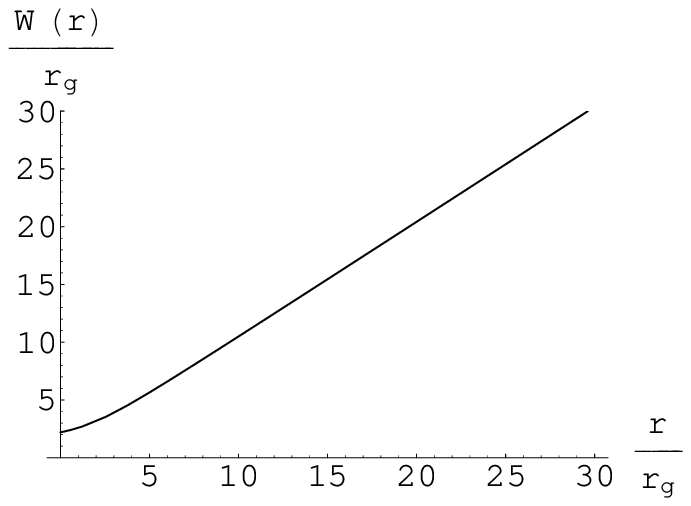}
\caption{Solution of the RTG equations for $\frac{r_g}{r_0} = 10^{-2}$.}
\label{rtg2}
\end{figure}

\begin{figure}
\epsfxsize=7cm\epsfysize=5cm\epsffile{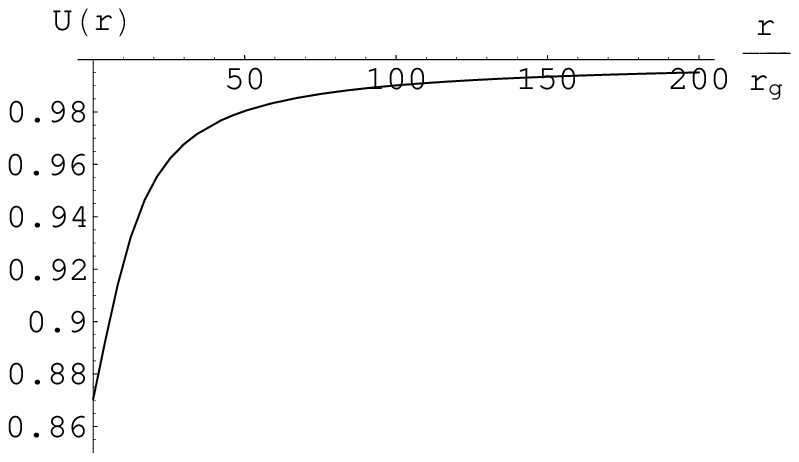}
\vskip -5cm
\hskip 9cm
\epsfxsize=7cm\epsfysize=5cm\epsffile{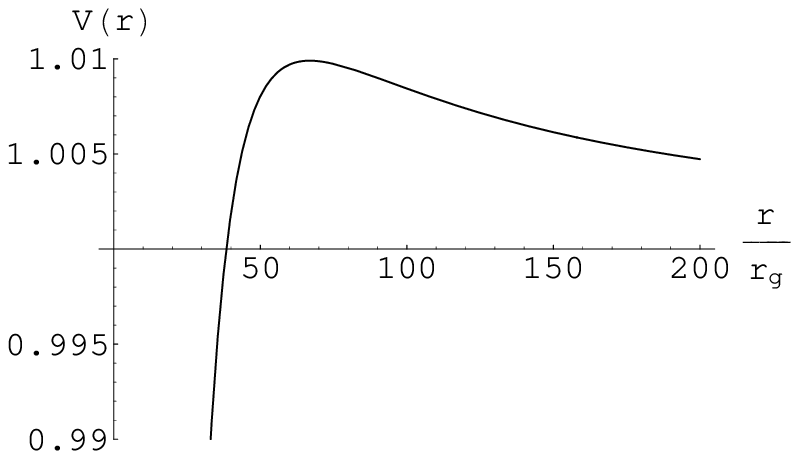}
\epsfxsize=7cm\epsfysize=5cm\epsffile{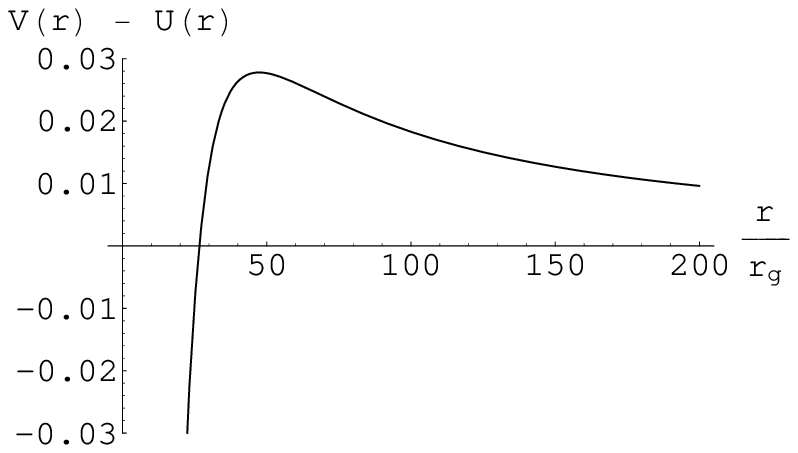}
\vskip -5cm
\hskip 9cm
\epsfxsize=7cm\epsfysize=5cm\epsffile{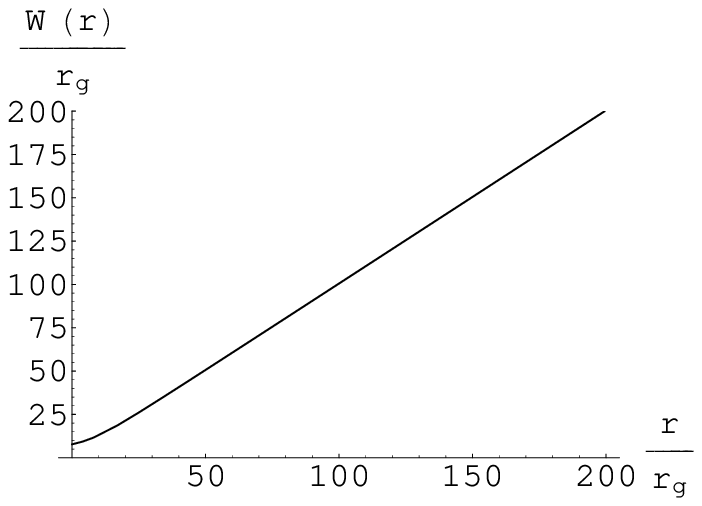}
\caption{Solution of the RTG equations for $\frac{r_g}{r_0} = 10^{-4}$.}
\label{rtg4}
\end{figure}

\begin{figure}
\epsfxsize=7cm\epsfysize=5cm\epsffile{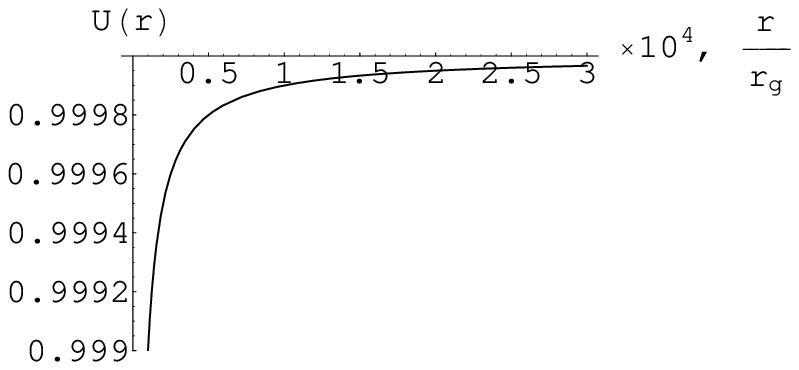}
\vskip -5cm
\hskip 8cm
\epsfxsize=7cm\epsfysize=5cm\epsffile{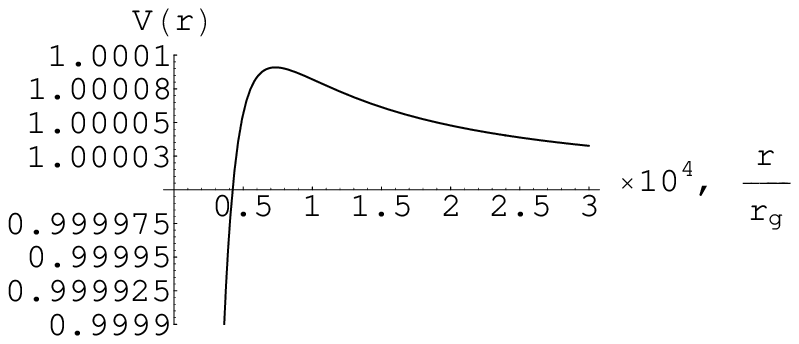}
\vskip -0.5cm
\epsfxsize=7cm\epsfysize=5cm\epsffile{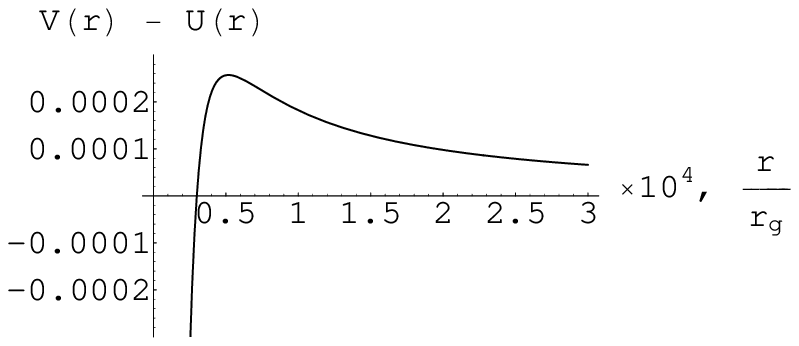}
\vskip -5cm
\hskip 8cm
\epsfxsize=7cm\epsfysize=5cm\epsffile{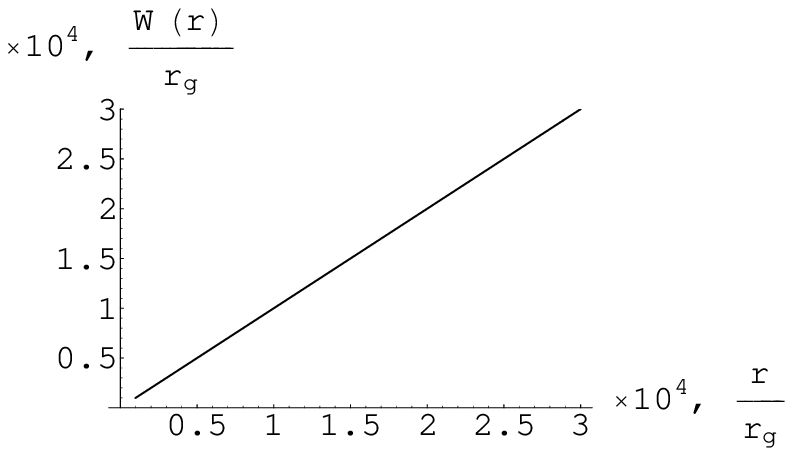}
\caption{Solution of the RTG equations for $\frac{r_g}{r_0} = 10^{-8}$.}
\label{rtg8}
\end{figure}

The RTG equations were integrated for $10^{-8}\le\frac{r_g}{r_0}\le 1$ in the 
interval $0\le r\le R$ where R had been chosen large enough for the asymptotic 
formulae (\ref{asymp}) to be valid. For some values of $\frac{r_g}{r_0}$
some fragments of the relevant solutions are represented in 
fig. (\ref{rtg2}) -- (\ref{rtg8}). These solutions are stable relative to the 
change of the integration accuracy or the choice of the starting point R. 
Let us enumerate their main features.
\begin{enumerate}
\item The RTG solution does not tend to the General Relativity solution 
at $r_0\to\infty$ (in other words, operations of solving the field equations 
and taking the limit $r_0\to\infty$ are not permutable).
\item All the metric coefficients in the solution are regular and not equal 
to zero everywhere in the considered region.
\item If $\frac{r_g}{r_0}<3\cdot 10^{-2}$ the solution has a crossover point 
$r_1$ such that $U(r)>V(r)$ at $r<r_1$. It means that the light cone of the 
effective curved space opens wider than the Minkowski space light cone 
in the region $r<r_1$ ($\lim_{r_0\to\infty}r_1 = \infty$). 
\item For any $\frac{r_g}{r_0}$ the effective metric coefficient $V(r)$ has a 
maximum at some point $r_2$, i.e. $V(r)$ demonstrates strongly 
non-Schwarzchild behavior at $r<r_2$ ($\lim_{r_0\to\infty}r_2 = \infty$).
\end{enumerate}
We managed to solve (\ref{rtgequ}) only for $10^{-8}\le\frac{r_g}{r_0}\le 1$. 
To find crucial points $r_1$ and $r_2$ of the solution for 
$\frac{r_g}{r_0}\le10^{-8}$ we used the rational interpolation. Now we can 
obtain some rough estimates of the value of the gravitational field mass if we 
assume some physical conditions as enumerated below:
\begin{enumerate}
\item The Mercury orbit must be located in the region $r>r_2$, $\Rightarrow$ 
$m_g>3\cdot 10^{-58}g$.
\item The light cone of the effective Riemann space for the case of neutron star 
must not open wider than the light cone of the Minkowski space, i.e. the neutron 
star radius  must be larger than $r_1$, $\Rightarrow$ $m_g>10^{-46}g$.
\end{enumerate}
Both the estimates of the lower limit for the gravitational field mass value are 
in strong contradiction with the upper limit $m_g<1.3\cdot 10^{-66}g$ obtained 
from the analysis of the Universe expansion~\cite{logunov}.\\

{\bf Acknowledgements:} I would like to thank A.A. Logunov, V.A. Petrov, 
V.V. Gusev and V.V.~Kiselev for stimulating discussions and useful criticism.

\vskip 2cm

\newpage


\begin{thebibliography}{99}

\bibitem{logunov} A.A. Logunov, {\bf The Theory of Gravity}. Moscow, NAUKA 2001.

\end{thebibliography}
\end{document}